\documentclass{IEEEtran}
\IEEEoverridecommandlockouts
\usepackage{cite}
\usepackage{amsmath,amssymb,amsfonts}
\usepackage{algorithmic}
\usepackage{graphicx}
\usepackage{textcomp}
\usepackage{comment}
\usepackage{times}
\usepackage{graphicx}
\usepackage{multirow}
\usepackage{url}
\usepackage[table,xcdraw]{xcolor}
\usepackage{subcaption}
\usepackage{cite}
\usepackage{anyfontsize}
\usepackage{caption}
\usepackage{authblk}
\usepackage[hidelinks,unicode]{hyperref}
\def\BibTeX{{\rm B\kern-.05em{\sc i\kern-.025em b}\kern-.08em
    T\kern-.1667em\lower.7ex\hbox{E}\kern-.125emX}}

\begin{document}

\title{HMM-based data augmentation for E2E systems for building conversational speech synthesis systems }
%\vspace{-1em}
\author{Ishika Gupta$^1$, Anusha Prakash$^2$, Jom Kuriakose$^1$, Hema A. Murthy}
% \author{Ishika Gupta$^1$, Anusha Prakash$^2$, Jom Kuriakose$^1$, Hema A. Murthy$^1$ \\ \email{ishika@cse.iitm.ac.in., anushaprakash@smail.iitm.ac.in \\ jom@cse.iitm.ac.in, hema@cse.iitm.ac.in}}
% \author[1]{Awais Adnan \\ \email{awaisadnan@gmail.com}}

\affil[1]{Department of Computer Science \& Engineering, Indian Institute of Technology Madras, India}
\affil[2]{Department of Electrical Engineering, Indian Institute of Technology Madras, India} 
% \email{ishika@cse.iitm.ac.in., anushaprakash@smail.iitm.ac.in, \\ jom@cse.iitm.ac.in, hema@cse.iitm.ac.in}
%{\footnotesize \textsuperscript{*}Note: Sub-titles are not captured in Xplore and
%should not be used}
%\thanks{Identify applicable funding agency here. If none, delete this.}
%\vspace{-6cm}
\maketitle
\begin{abstract}
This paper proposes an approach to build a high-quality text-to-speech (TTS) system for technical domains using data augmentation. An end-to-end (E2E) system is trained on hidden Markov model (HMM) based synthesized speech and further fine-tuned with studio-recorded TTS data to improve the timbre of the synthesized voice. The motivation behind the work is that issues of word skips and repetitions are usually absent in HMM systems due to their ability to model the duration distribution of phonemes accurately. 
%Tree-based clustering and state-tying take care of the unseen context using a question set. 
Context-dependent pentaphone modeling, along with tree-based clustering and state-tying, takes care of unseen context and out-of-vocabulary words.
% Moreover, it can scale for out-of-vocabulary words since HMM uses context-dependent pentaphones 
% Moreover, it can scale for out-of-vocabulary words since HMM uses pentaphone context. 
A language model is also employed to reduce synthesis errors further.

Subjective evaluations indicate that speech produced using the proposed system is superior to the baseline E2E synthesis approach in terms of intelligibility when combining complementing attributes from HMM and E2E frameworks. The further analysis highlights the proposed approach's efficacy in low-resource scenarios.
\end{abstract}


\section{Introduction}
With advancements in text-to-speech (TTS) synthesis techniques, rapid progress has been made in producing both intelligible and natural speech. 
%Although the speech quality is very close to the human voice, the generated voice lacks resilience to the errors afflicting the speech output due to skipped words, repetitions, and babbling issues. 
E2E systems\footnote{In this paper, we consider the E2E systems having a 2-stage pipeline, i.e., mel-spectrogram generation followed by audio synthesis using a vocoder.} that are trained on grammatically correct read speech do not scale for conversational speech synthesis.  Building conversational speech synthesis systems is difficult owing to the presence of disfluencies, significant syllable rate variation, and prosody variation.
\cite{comini2022low}, \cite{o2022combining} are some recent works that highlight the challenges in the context of conversational speech synthesis. %{\textbf{Note: Ishika -- please refer to the conversational speech synthesis systems in the literature here.  I think there were a few in IS 2022} } (addressed)
%There are more than 7000 languages spoken in the world and most of the languages do not have enough training speech dataset. Hence, building a good quality TTS system in a low-resource scenario is challenging for such languages. 
One such scenario where TTS accuracy is important is in the task of transcreating English educational classroom lectures from National Programme on Technology Enhanced Learning (NPTEL) \cite{NPTEL} into other languages where having an error-free TTS system is essential. In applications like this, even a small error in the audio can degrade the listener's experience. The main objective of this paper is to develop a good quality TTS system for reproducing technical videos in multiple Indian languages where the video is kept as it is while the content is translated and regenerated in Indian languages and integrated with the original video. The source videos are technical lectures and use a style of speech that is conversational. E2E systems do not scale for conversational speech, as sentences are seldom grammatically correct.
%A speech-to-speech transcreation (S2ST) system replaces the original speech content contained in a video with a speech in a different language such that it matches the naturalness of the original video \cite{m21_ssw}. The S2ST module can be divided into 4 phases: (i) transcribing text content from a speech in a source language (English) using an automatic speech recognizer (ASR), (ii) translating text from a source to a target language (machine translation- MT), (iii) synthesizing speech from the translated text (TTS), and (iv) lip-syncing of synthesized speech with the original video. 
%The ASR output in English is generated in the SubRip Text (SRT) format. This is especially true for English to Indian language translation, where the word order in English and Indian languages can be significantly different\cite{subbarao2008typological}.
An initial experiment is performed, wherein the translated text, conversational in nature is synthesized using a Tacotron2-based E2E system trained on read-speech data. Conversational text is characterized by an unplanned set of words and  abrupt sentence ending, which may also contain disfluencies like ` umm'
, `hmm' etc. It is observed that the synthesized output is often prone to errors. This may be primarily due to the following reasons: (i) TTSes are generally trained using read speech, and (ii) the contexts seen in test data are seldom seen in train data.
\vspace{-0.2cm}
\begin{figure}[!ht]
  \centering
  \captionof{table}{Conversational text structure in different scenarios}
  \label{fig:conversational_txt}
  \includegraphics[width=\linewidth]{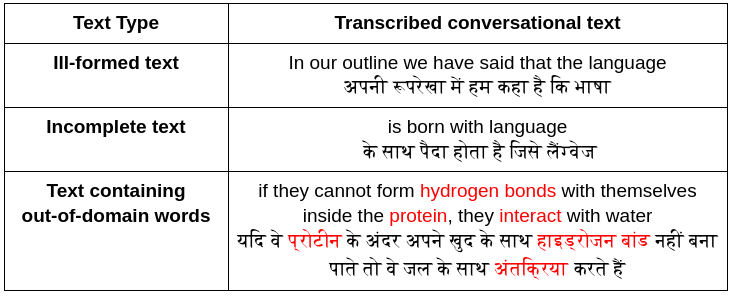}
\end{figure}
\vspace{-0.2cm}
Table \ref{fig:conversational_txt} shows different types of issues in Hindi text construction frequently encountered in the translated text.
%Their corresponding original text in English has also been provided for reference. Although the English text is syntactically correct in the first example, its translated Hindi text is not. 
Hence the synthesis of conversational text taken from technical lectures becomes challenging. In this paper, we propose a novel data augmentation approach where synthetic utterances generated from a statistical HMM-based speech synthesis system (HTS) \cite{zen2007hmm} are used in tandem with a language model. This gives us the added advantage of not only increasing labeled context-rich training data but also helping in broadening the domain of the TTS system. Data augmentation is a popular technique that has been used in many research areas to help complement inadequacies in the training data \cite{zheng2021using, tsunoo2021data, mikolajczyk2018data, perez2017effectiveness}. In the current work, we synthesize a large amount of multi-domain text in the HTS framework and augment it with the original $<$text, read speech audio$>$. Owing to state clustering in HTS systems, HTS degrade gracefully for unseen contexts.
%The motivation for using the HMM-based framework is its robustness in synthesizing text from any domain.
E2E synthesis is prone to errors, such as word skips, repetitions, and substitutions \cite{li2021speech, yu2020durian}, especially for conversational text. 
%The HTS framework models context-dependent pentaphones, and trains separate acoustic and duration models. 
%Tree-based clustering and explicit duration modeling aid in producing almost error-free synthesis, albeit with reduced speech quality. 
%An E2E model is first trained on HTS-synthesised speech and is further fine-tuned on studio-recorded data to improve the timbre of synthesized speech. The idea is to exploit the best of both paradigms: robustness from HTS and synthesis quality from neural-based E2E TTS systems.
%A language modeling-based technique is also proposed to further reduce errors in synthesis.

As a part of this work, HTS and E2E voices are trained for two Indian languages-- Hindi and Tamil. For the E2E paradigm, the autoregressive (AR) Tacotron2 architecture and Waveglow vocoder are used. For comparison, systems are also built by augmenting data using E2E (Tacotron 2) synthesized utterances, which is then fine-tuned on clean TTS data. The systems are analyzed based on the errors made on a conversational text, and comprehension mean opinion score (CMOS) and subjective evaluations are also conducted. 
Experimental results indicate that the proposed TTSes in both languages achieved an average CMOS of 94.41\% and an average MOS score of 3.97.

%We also study the performance of the various systems in a low-resource scenario. A non-auto-regressive (NAR) FastSpeech TTS system \cite{ren2019fastspeech} is also included for comparison in the experiments section. We show that the FastSpeech system trained on low amounts of data do not perform well, whereas the proposed system can prove its usefulness in such a case.
%{\bf Note: Ishika Fastspeech work can go to the ICASSP Paper} (addresed)
\par The rest of the paper is organized as follows. Section \ref{sec:related_work} reviews the related literature. The proposed approach is presented in Section \ref{sec:proposed_system}. Section \ref{sec:experiments} details the experiments conducted and achieved results. Finally, Section \ref{sec:conclusion} concludes the paper highlighting open questions for future research.
%\vspace{-0.1cm}
\section{Related Work}
\label{sec:related_work}
%\vspace{-1em}
%Deep learning-based state-of-the-art TTS models rely on extensive training data sets. Since a large amount of labeled speech data is difficult to acquire for Indian languages, improving TTS systems under low resource conditions has become topical.
To address the issue of data scarcity among low-resource languages, several methods have been developed using synthetic audio datasets, including data augmentation  \cite{zheng2021using, hwang2021tts, huybrechts2021low} and domain adaptation \cite{chu2002domain}. 

%{\bf Note: Ishika Is this done for Indian languages?} (not done for Indian Languages)
%In \cite{zheng2021using}, synthetic data is used for training automatic speech recognizers (ASR) to boost the recognition of out-of-vocabulary (OOV) words. 
In TTS applications, \cite{hwang2021tts} generates a large amount of synthetic utterances using an auto-regressive (AR) model to improve the quality of non-AR TTS models. \cite{huybrechts2021low} uses synthetic data in the desired speaking style on top of the available recordings generated by applying voice conversion. Further fine-tuning the model on a small amount of expressive samples for the target speaker helps improve naturalness and style adequacy. \cite{chu2002domain} focuses on increasing the naturalness of TTS systems on specific domains by adding domain-specific speech to the unit inventory.
 
Parallel efforts to use hybrid paradigms for TTS have also been explored. \cite{mano2020hybrid} uses a hybrid approach combining HTS with the neural-network-based WaveGlow vocoder \cite{prenger2019waveglow} using histogram equalization (HEQ) in a low resource setting for improving the quality of speech output. Attempts have also been made by incorporating bidirectional encoder representations from transformers (BERT) pre-trained on large amount of text corpus in the recurrent neural networks-based speech synthesis framework and fine-tuning the system on speech domain to enhance the prosody and pronunciation of generated speech \cite{kenter2020improving, jia2021png}.
 But none of these approaches address the challenge of synthesizing conversational type text, which may contain out-of-domain words. In our proposed work, we generate synthetic data in the HTS framework, which is more robust than using E2E augmented data employed in \cite{hwang2021tts}. To the best of our knowledge, the proposed method is the first to combine the best features of two different TTS frameworks in tandem with an external language model to synthesize conversational text. Apart from the autoregressive models, the state-of-the-art non-autoregressive models FastSpeech and FastSpeech2 \cite{ren2020fastspeech} also nearly eliminate the problem of word skipping and repeating in particularly hard cases but at the expense of good duration alignments obtained by autoregressive teacher model trained on a large dataset. 
 
\section{Proposed Approach}
\label{sec:proposed_system}
\begin{figure}[!ht]
  \centering
  \includegraphics[width=\linewidth, height=8cm]{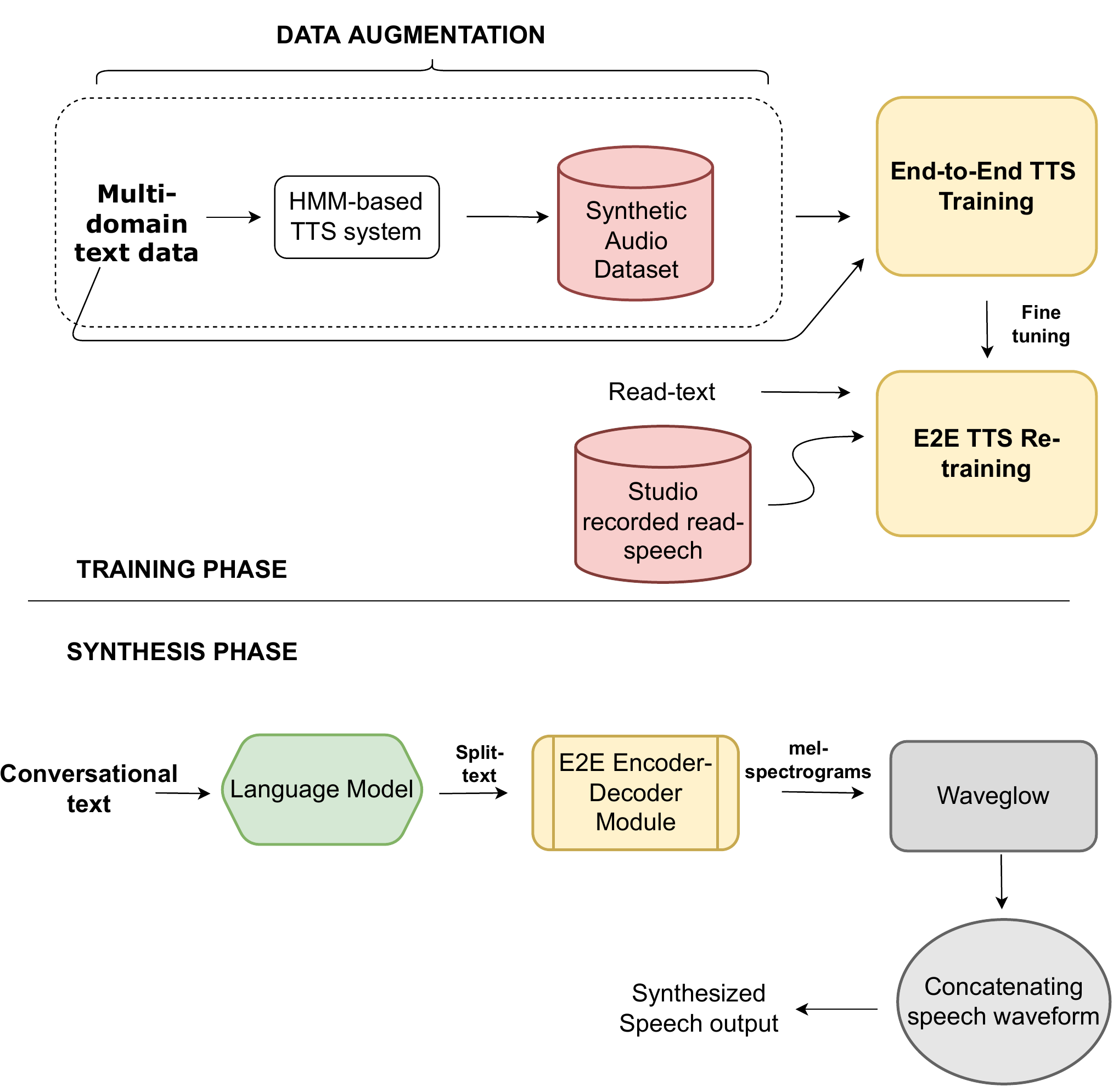}
  %\vspace{-0.7cm}
  \caption{{\fontsize{8}{10}\fontfamily{ptm}\selectfont Training and synthesis phases of the proposed approach}}
  \label{fig:training}
  
\end{figure}

The training framework of the proposed system is depicted in Fig. \ref{fig:training}. The approach is divided into training and synthesis phases. In the training phase, we first train an HTS model with studio-recorded read speech data. Then we synthesize a large amount of multi-domain text in the HTS framework. 
%{\bf Ishika -- you must compare and contrast TTS data and the generated text}
With the generated $<$text, audio$>$ pairs, an E2E model is trained. Since the voice quality of HTS-synthesized utterances is more robotic in nature, this effect is also seen in the synthesis of the resultant E2E model. Hence, this model is fine-tuned on the same studio-recorded data to restore the timbre of the speaker. In conventional synthesis, the text is directly given to the final E2E model to generate mel-spectrogram frames. Then speech is reconstructed from the mel-spectrogram features using a WaveGlow vocoder. Waveglow is a non-autoregressive flow-based network capable of generating high-quality speech from mel-spectrograms. It is to be noted that the WaveGlow model is also trained on the same studio-recorded data.
 
 In some instances, errors are present in synthesis mainly due to the grammatically incorrect phrases/sentences, further compounded by the auto-regressive nature of the E2E Tacotron2-based model. To circumvent this, a language model is employed to split the text based on a drop in the log-likelihood scores. 
\vspace{-0.5cm}
\subsection{Multi-domain text data collection}
We crawled text from Kaggle \cite{GTrans} web source in each language for augmenting data. The text data was carefully collected from various domains ranging from health, lifestyle, entertainment, science, technology, and ordinary colloquial conversations to WH questions and Yes-No questions. The intent of having such diversity in text data is to train the TTS model with a lot of context information and enable the TTS model to learn various kinds of phonotactics.
To prepare the training set text scripts, the collected text data is first cleaned and curated by removing special symbols, and text normalization is performed to achieve data uniformity. This step included transliterating numbers, English words, and acronyms in the native language. Totally we have selected around 14,450 and 11,043 sentences ($\approx$ 15 hours) in Hindi and Tamil, respectively.

\subsection{HTS training and its importance}
To train the HTS model, the text is first transcribed in terms of its constituent phones which are represented using the common label set (CLS) representation \cite{ramani2013common}. Then the speech waveform is aligned at the phone level using a hybrid HMM-DNN approach. Using the aligned speech data, the HTS voice is trained on context-dependent pentaphone units. Two models are trained -- the duration model and the acoustic model. 
%The text is converted into its constituent phones and presented to the HTS voice at synthesis. 
Phone HMMs are concatenated to obtain the sentence HMM \cite{baby2020significance}.
%{\bf cite the hybrid paper -- Arun Baby here}
The duration model predicts the number of frames reserved for each phone. The acoustic model predicts the acoustic features for the required number of frames. The HTS engine internally uses a Mel-generalized log spectrum approximation (MLSA) vocoder,  which is used to synthesize the utterances. The explicit modeling of a duration predictor corresponding to context-dependent pentaphones ensures that the synthesis is robust and intelligible. Hence, compared to the auto-regressive E2E model, the HTS model makes fewer mistakes during synthesis.

\begin{figure}[!ht]
  \centering
  \includegraphics[width=0.95\linewidth]{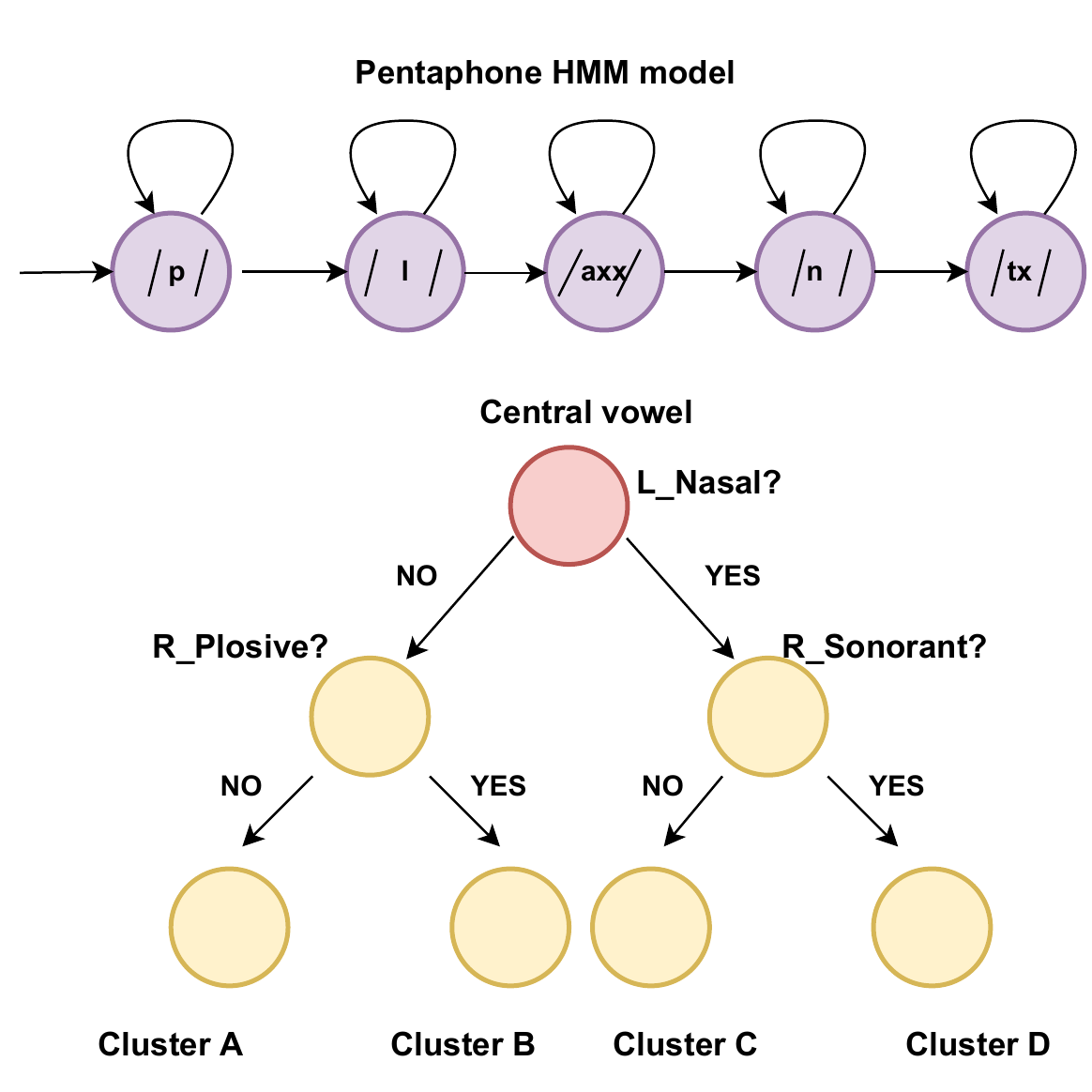}
  \caption{{\fontsize{8}{10}\selectfont Example of a decision tree-based clustering for a context dependent pentaphone HMM model. ``R'' and ``L'' refer to the right and left contexts, respectively.}}
  \label{fig:tree-based_clustering}
%\vspace{-0.4cm}
\end{figure}

A decision tree-based clustering is performed during HTS training to take care of unseen context based on a phonetic yes-no question set, which is frequently encountered in the technical domain \cite{wu2016merlin}. It follows a top-down approach which uses phonetic knowledge together with the training data to decide which contexts are acoustically similar. After the clustering, the context-dependent pentaphone models, which are acoustically similar, share the same parameters and HMM states at the leaf nodes. An example of tree-based clustering is illustrated in Fig. \ref{fig:tree-based_clustering}. This technique ensures that the HMM states are also shared with the unseen phoneme context. Thus, HTS can broadly extrapolate to out-of-domain word scenarios. However, the HTS synthesized utterances have muffled quality due to averaging of model parameters that occurs at the leaf nodes of the decision tree.
%\vspace{-0.4cm}

%The current work employs a hybrid (group delay) GD-DNN segmentation algorithm proposed in \cite{baby2017deep} for training the HTS system. In \cite{baby2017deep}, boundaries generated by DNNs are corrected using signal processing-based GD cues. \cite{baby2017deep} also highlights the importance of training systems using accurate boundaries. Hence, the explicit duration modeling in HTS ensures no word skips and repetitions in synthesis, which is not the case in E2E Tacotron2 based synthesis.
%\vspace{-0.2cm}
\subsection{E2E system training}
In our work, we have considered the Tacotron2 architecture \cite{shen2017natural} as the E2E framework. It employs a sequence-to-sequence encoder-decoder-based architecture with a soft attention mechanism. It takes in a sequence of characters or phonemes as input and predicts mel-spectrograms frame by frame in an autoregressive manner. The unified parser for Indian languages is used to obtain phone-level transcriptions from grapheme-based text \cite{baby2016unified}. Finally, raw acoustic waveforms are obtained using a non-autoregressive WaveGlow vocoder.
For adaptation, the model parameters of the E2E model trained on synthetic data are fine-tuned on the clean studio-recorded TTS data for further 200 epochs. 
The stopping criteria used for the experiments ensure that the attention weights are approximately diagonal.
%This step is mainly performed to improve the timbre of the synthesized speech.
\subsection{Language Modeling}
%\vspace{-0.4cm}
\begin{figure}[!ht]
  \centering
  \includegraphics[width=0.95\linewidth, height=5cm]{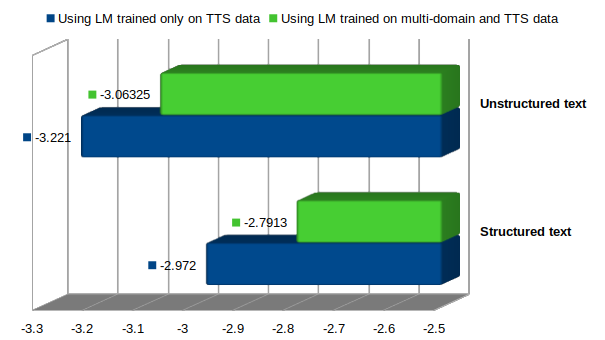}
  \caption{{\fontsize{8}{10}\selectfont Average log likelihood scores on structured and unstructured transcription from a NPTEL lecture}}
  \label{fig:avg_likelihood}
\vspace{-0.3cm}
\end{figure}
%\vspace{-0.5cm}

%\vspace{-0.3cm}
A language model (LM) is employed to further reduce the errors in synthesis. We have used the SRILM toolkit \cite{stolcke2002srilm} for building a statistical LM in our work. The LM estimates the relative likelihood of the sequence of words in a sentence. A bigram LM is first trained on the training text (TTS text or TTS text$+$multi-domain text). We use Kneser-Ney \cite{zhang2014kneser} as a smoothing technique to account for unseen bi-gram pairs. We obtain a log-likelihood score corresponding to every bi-gram pair for the text. If the score for any bi-gram pair is less than a particular threshold, the next word is less likely to occur after the previous one. Hence, we split the text at that point, thereby reducing the poor grammatical structure to some extent. Then the smaller segments are synthesized and finally concatenated together.

%\vspace{-0.3cm}
\begin{figure}[!ht]
  \centering
  \captionof{table}{Example of a text being split using a language model}
  \label{fig:split-text}
  \vspace{-0.3cm}
  \includegraphics[width=\linewidth]{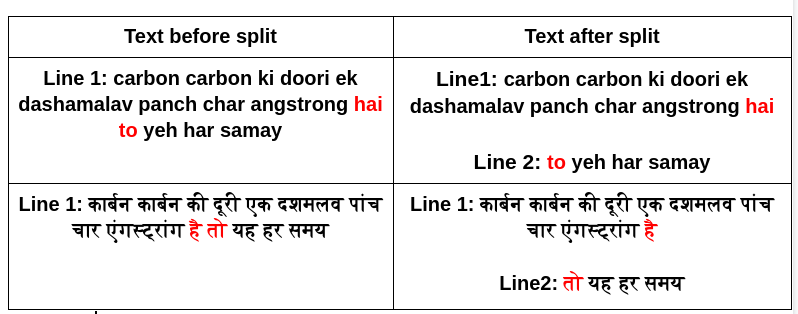}
\end{figure}
%\vspace{-0.3cm}
Table \ref{fig:split-text} shows the split in test sentence as the probability of $'/to/'$ occuring after $'/hai/'$ is low, hence the split happens after $'/hai/'$.
The threshold value can vary depending on the domain of training data used, so this value is chosen empirically.   We have used the log-likelihood score of $- 5.12$ as the threshold for our experiments.

\begin{figure}[!ht]
  \centering
  \includegraphics[width=0.95\linewidth, height=5cm]{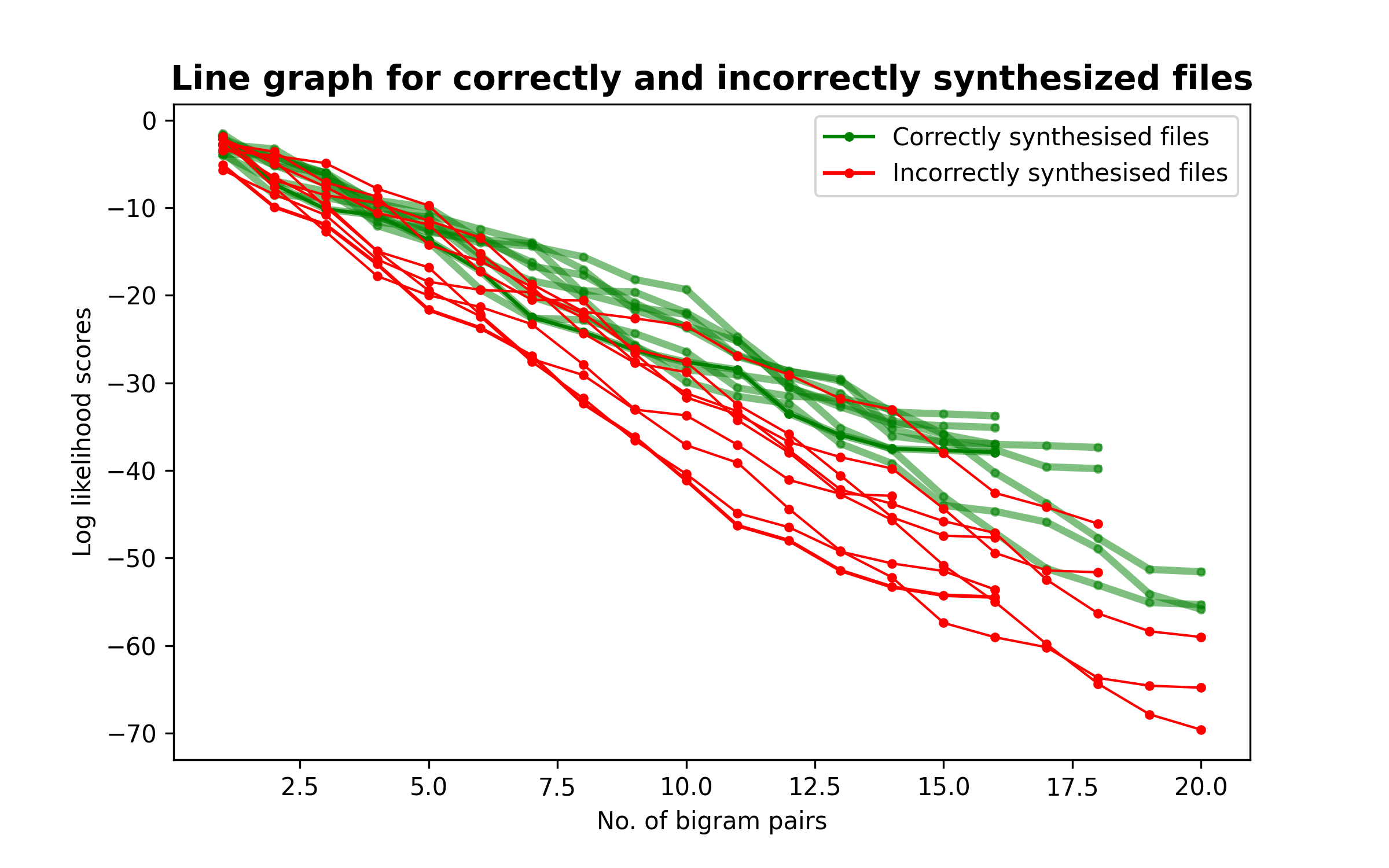}
  \caption{{\fontsize{8}{10}\selectfont Log-likelihood scores of each bigram pair for structured and unstructured sentences}}
  \label{fig:line_graph}
  \vspace{-0.3cm}
\end{figure}

Fig. \ref{fig:avg_likelihood} shows the average log-likelihood scores of structured and unstructured text corresponding to a technical NPTEL lecture. These scores are normalized by the number of words in each text. It is seen that the log-likelihood scores are higher for structured text compared to unstructured text. We also see that LM trained with multi-domain$+$TTS text has higher log-likelihood scores, indicating that the augmented TTS system captures more context. Moreover, Fig. \ref{fig:line_graph} supports that the cumulative log-likelihood score in a structured sentence is greater than that of unstructured sentences of the same length.
Language model-based splitting of text is used as a preprocessor before the text is given to the E2E system for synthesis.
This reduces issues in the synthesized speech due to the text's ill-formed structure, mainly observed in conversational-type transcriptions.
\vspace{-0.3cm}
\vspace{-0.2cm}
\section{Experiments and Results}
\label{sec:experiments}
Experiments are conducted in two Indian languages-- Hindi and Tamil. Studio recorded Hindi male (8.5 hours) and Tamil male (10 hours) datasets from the open-source IndicTTS database \cite{baby2016resources} are considered. 
%\vspace{-2mm}
%We also remove utterances greater than 15 seconds to avoid poor attention alignments that may occur with long audio files. 
We use the HTK \cite{HTS_toolkit} \cite{young1993htk} and ESPnet toolkits \cite{watanabe2018espnet} for building voices with default configurations. A pre-trained Waveglow model of the LJ Speech dataset is fine-tuned on 8.5 hours of read speech data for each language. Table \ref{tab:systems} summarizes the various systems trained for each dataset. The text in parentheses indicates the training data used. In this work, we have considered two baseline systems. System 2 trained on 8.5 hours of speech is Baseline 1, and System 4 trained on 15 hours (E2E synthesized speech) + 8.5 hours (original read speech) of data is Baseline 2 as indicated in Table \ref{tab:systems}. The proposed TTS system and Baseline 2 TTS are trained on equal amounts of data for a uniform comparison.

The experiments have been conducted by considering the Tacotron2 model as the E2E model. 
%Only the low resource scenario experiments have been carried out with Tacotron2-based and FastSpeech-based E2E models. 
%\vspace{-0.2cm}

\begin{table}[h!]
\centering
\caption{List of systems trained}
\label{tab:systems}
\begin{tabular}{|c|c|l}
\cline{1-2}
\textbf{Tag}                                                    & \textbf{TTS Model}                                                                                            &  \\ \cline{1-2}
System 1                                                        & HTS (read speech)                                                                                             &  \\ \cline{1-2}
\begin{tabular}[c]{@{}c@{}}System 2\\ (Baseline 1)\end{tabular} & E2E (read speech)                                                                                             &  \\ \cline{1-2}
System 3                                                        & E2E (E2E synthesised multi-domain text)                                                                       &  \\ \cline{1-2}
\begin{tabular}[c]{@{}c@{}}System 4\\ (Baseline 2)\end{tabular} & \begin{tabular}[c]{@{}c@{}}E2E (E2E synthesised multi-domain text) \\ +fine-tuned (read speech)\end{tabular}  &  \\ \cline{1-2}
System 5                                                        & E2E (HTS synthesised multi-domain text)                                                                       &  \\ \cline{1-2}
System 6                      (Proposed)                                  & \begin{tabular}[c]{@{}c@{}}E2E (HTS synthesized multi-domain text) \\  +fine-tuned (read speech)\end{tabular} &  \\ \cline{1-2}
\end{tabular}
\end{table}   
   
\begin{figure}[t]
  \centering
  \includegraphics[width=\linewidth, height=5cm]{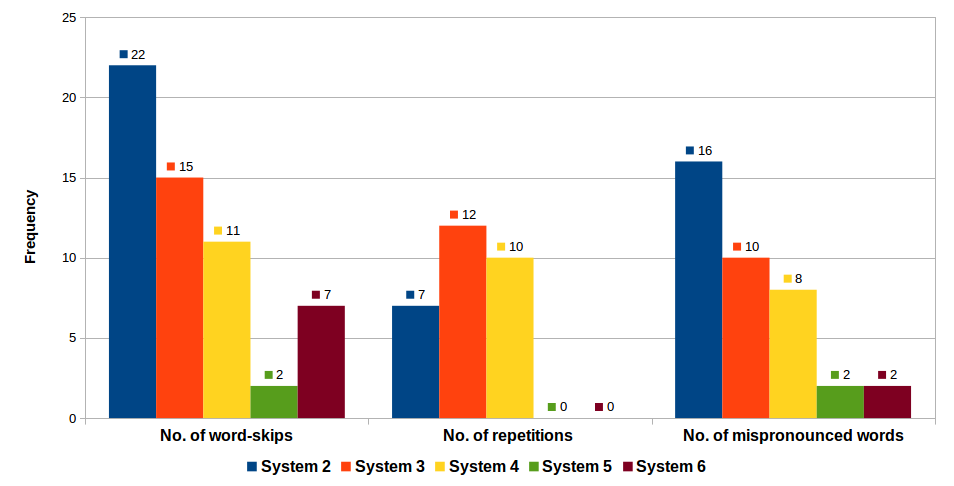}
  \caption{{\fontsize{8}{10}\selectfont Error statistics of a Hindi technical lecture (without language model)}}
  \label{fig:error-statistics_hindi}
\end{figure}
      
\begin{figure}[t]
  \centering
  \includegraphics[width=\linewidth, height=5cm]{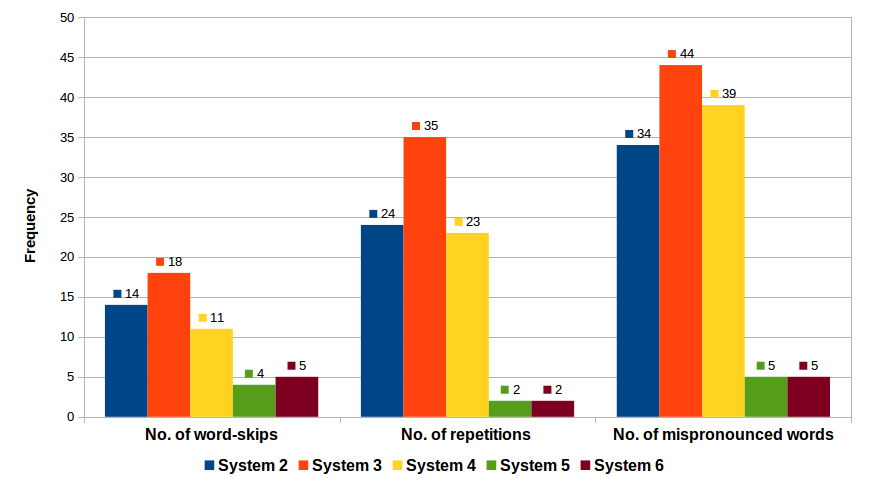}
  \caption{{\fontsize{8}{10}\selectfont Error statistics of a Tamil technical lecture (without language model)}}
  \label{fig:error-statistics_tamil}
 % \vspace{-0.6cm}
\end{figure}
\vspace{-0.2cm}
\subsection{Analysis of synthesized speech}
\begin{figure}[t]
  \centering
  \includegraphics[width=\linewidth, height=5cm]{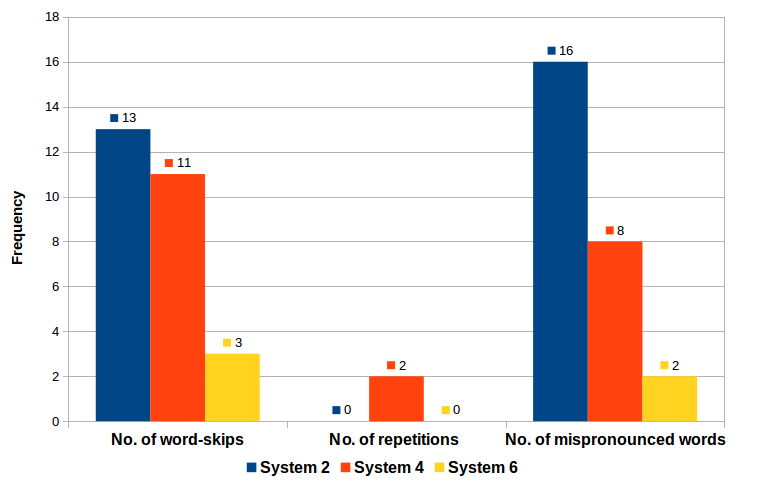}
  \caption{{\fontsize{8}{10}\selectfont Error statistics of a technical lecture in Hindi (with language model)}}
  \label{fig:error-statistics_lm}
\end{figure}

 A set of Swayam lectures \cite{paul2018swayam}, comprising 347 (for Hindi) and 350 sentences (for Tamil), has been considered for calculating error statistics. Fig. \ref{fig:error-statistics_hindi} and \ref{fig:error-statistics_tamil} present the error statistics across various systems (without language model). Three types of errors are considered for analysis -- word skips, repetitions, and mispronunciations. These errors are counted manually from the Swayam lecture considered. It is seen that the errors made by HTS-augmented systems (Systems 5 and 6) are significantly less compared to System 2 (baseline) and E2E-augmented systems (Systems 3 and 4). A small increase in the number of word skips for System 6 can be attributed to the mismatch in the acoustic characteristics of original and synthetic training data. A similar trend is also observed with the language model (Fig. \ref{fig:error-statistics_lm}). Comparing Fig. \ref{fig:error-statistics_hindi} and \ref{fig:error-statistics_lm}, we see that the errors are reduced further when language model is used.

%  We also show pitch variations observed in the case of the proposed system compared to the baseline system (Figure \ref{fig:pitch_contour}). We see that the pitch contour of the word $`/mehetwapoorna/'$ is relatively flat for the E2E system (middle panel) compared to the HTS (top panel) or proposed system (bottom panel). This indicates that the proposed system can produce prosodically richer synthesis.
 
% \vspace{-0.2cm}
\begin{table}[!ht]
\centering
\caption{Comprehensibility MOS scores for Hindi TTS}
\label{tab:comprehensibility_test_hi}
\begin{tabular}{|c|c|c|c|}
\hline
\textbf{\begin{tabular}[c]{@{}c@{}}Systems \\ Evaluated\end{tabular}} & \textbf{\begin{tabular}[c]{@{}c@{}}MOS\\ Score\end{tabular}} & \textbf{Intelligibility} & \textbf{\begin{tabular}[c]{@{}c@{}}Comprehension\\ Score\end{tabular}} \\ \hline
Baseline 1 TTS                                                            & 3.486                                                        & 4.241                    & 94.25\%                                                                \\ \hline
Proposed TTS                                                              & 4.379                                                        & 4.612                    & 95.98\%                                                                \\ \hline
\end{tabular}
\end{table}
\vspace{-0.2cm}
\begin{table}[!ht]
\centering
\caption{Comprehensibility MOS scores for Tamil TTS}
\label{tab:comprehensibility_test_tam}
\begin{tabular}{|c|c|c|c|}
\hline
\textbf{\begin{tabular}[c]{@{}c@{}}Systems \\ Evaluated\end{tabular}} & \textbf{\begin{tabular}[c]{@{}c@{}}MOS\\ Score\end{tabular}} & \textbf{Intelligibility} & \textbf{\begin{tabular}[c]{@{}c@{}}Comprehension\\ Score\end{tabular}} \\ \hline
Baseline 1 TTS                                                            & 2.903                                                        & 3.660                 & 79.16\%                                                                \\ \hline
Proposed TTS                                                              & 3.570                                                        &  4.232                   & 92.85\%                                                                \\ \hline
\end{tabular}
\end{table}

%\vspace{-0.5cm}
%\vspace{-0.2cm}
% \begin{figure}[t]
%   \centering
%   \includegraphics[width=0.75\linewidth, height=6cm]{LaTeX/pitch.png}
%   \caption{Comparing variation of pitch throughout the utterance across HMM-based, E2E baseline, and proposed TTS systems respectively}
%   \label{fig:pitch_contour}
%   \vspace{-0.4cm}
%  % \vspace{-0.5cm}
% \end{figure}   
%\vspace{-0.2cm}
% \subsubsection{Qualitative Analysis}
% Essentially there are 2 types of evaluation tests that are used to measure the performance of TTS System.
%\vspace{-0.4cm}
\subsubsection{Subjective listening tests}
We perform subjective evaluations for comprehension sentences from multiple domains across the baseline and proposed systems. The point to note here is that we have considered only Baseline 1 in the evaluation test. We noticed that Baseline 2 (System 4) performed poorly than Baseline 1 as it was trained on synthetic speech generated by Tacotron 2 E2E model, which may have induced word skips or repetitions in the synthetic audio data.
%For example, Tacotron2 model trained with read speech and tested with conversational speech is prone to have more errors. 
The comprehension mean opinion score (CMOS) test is performed for Hindi and Tamil to assess the system performance and naturalness of the voice for Baseline 1 and the proposed system. It is a subjective evaluation test that not only assesses the quality of the speech but also checks whether the listener is able to comprehend the content of the audio clearly.
%{\bf Ishika: Please refer to the paper that discusses this} 
CMOS score are presented in Table \ref{tab:comprehensibility_test_hi} and Table \ref{tab:comprehensibility_test_tam} for Hindi and Tamil, respectively. 

Since the content is technical lectures, intelligibility is crucial.   A topic of interest is synthesized using various systems. Listeners are then quizzed to check the following: a) the audio quality is not monotonous, that the listener does not switch off, and b) that the content is clearly articulated.  
The test includes questions based on the audio content. 25 native evaluators of Hindi and 20 native evaluators of Tamil were asked to listen and rate 4 paragraphs presented in random order. The focus of the work is to see if the proposed system can improve improperly synthesized utterances while maintaining their naturalness. Hence such utterances are used in the evaluation test. MOS is calculated based on the audio quality rating. Listeners were asked to rate the quality of synthesized speech on a scale ranging from 1-5, with 1 being poor and 5 being human-like. Intelligibility is calculated based on the listener's understandability score on a similar scale of 1-5. For the comprehension test, 4 paragraph questions with a mix of technical lectures and stories were considered. Based on the content of the audio clip, basic questions were asked related to the audio content to see how well the native users could understand and perceive it. The comprehension score for each system is obtained by calculating the percentage of correct answers in the evaluation test. We see that the MOS, intelligibility, and comprehension scores are better for the proposed system than the baseline system.
To make the results more concrete, we have also performed a p-test to prove the statistical significance of comprehensive MOS and intelligibility scores mentioned in Tables \ref{tab:comprehensibility_test_hi} and  \ref{tab:comprehensibility_test_tam}. The p-values are reported in Table \ref{tab:statistical_significance}. Both the p-values are less than 0.05. Hence the MOS scores are statistically significant.
%\vspace{-0.3cm}
\begin{table}[!ht]
\centering
\caption{Statistical significance test for CMOS scores based on voice quality and intelligibility}
\label{tab:statistical_significance}
\begin{tabular}{|c|c|}
\hline
\textbf{Statistical significance for} &  \textbf{p-value} \\ \hline
CMOS score                                            & 0.0001                                    \\ \hline
Intelligibility                                         & 0.019      \\ \hline
\end{tabular}
 \end{table}
%\vspace{-0.5cm}
\vspace{-0.3cm}
\subsection{Analysis of TTS systems in a low-resource scenario}
In this section, we have conducted experiments for Hindi to evaluate our proposed TTS system in a low-resource setting and compare it against the E2E Tacotron2-based and non-autoregressive FastSpeech model. For this experiment, 1 hour of $<$text, speech$>$ data is used, which is a subset of the studio-recorded data used in the previous experiments. 

The proposed system is trained on the multi-domain HTS synthesized audio and then fine-tuned on this 1 hour of data (System 6-1hr). It is to be noted that the HTS system is also trained with 1 hour of data. The Tacotron2-based E2E model is trained only with 1 hour of TTS data (System 2-1hr). To train the FastSpeech model in an E2E manner, the Tacotron2 model (System 2-1hr) is considered as a teacher model. For the low-resource training data, alignments are predicted by the Tacotron2 model in teacher-forcing mode. These alignments are then fed to the FastSpeech model for training.

Among the three systems, it is observed that both E2E Tacotron2 and FastSpeech models failed to train on 1 hour of data and performed poorly. The models generated uncomprehending gibberish speech. Being a statistical parametric model, since the HTS system is able to train well even with low training data \cite{pine2022requirements}, the proposed system performs reasonably well over the Tacotron2-based and FastSpeech models. We also compare the performance of these 3 systems for Hindi by conducting a degradation mean opinion score (DMOS) test. The DMOS test is similar to the MOS test \cite{viswanathan2005measuring}, except that the score of each system is normalized with respect to that of the original ground truth audio. 15 native listeners in Hindi were asked to listen to a set of 22 utterances (6 for each system + 4 natural) and rate it on the basis of intelligibility for 3 systems. The utterances taken for evaluation are completely unseen test sentences. Table \ref{tab:dmos} presents the DMOS scores for the systems under evaluation. The DMOS scores for E2E Tacotron2-based and FastSpeech were close to 1, so we didn't report the values in the table.
\begin{table}[h!]
\centering
\caption{DMOS scores of Hindi TTS systems in a low resource scenario (studio-recorded training data: 1 hour)}
\label{tab:dmos}
\begin{tabular}{|c|c|l}
\cline{1-2}
\textbf{Systems Evaluated} & \textbf{DMOS/Speech Output} &  \\ \cline{1-2}
Proposed model (System 6-1hr)  & 3.04             &  \\ \cline{1-2}
Tacotron-based E2E (System 2-1hr)        & gibberish               &  \\ \cline{1-2}
FastSpeech-based E2E                & gibberish               &  \\ \cline{1-2}
\end{tabular}
\vspace{-0.1cm}
\end{table}

 It is seen that the performance of the proposed approach is superior to that of the end-to-end systems. %Hence, systems trained completely in an E2E manner for low resource scenarios do not perform well.
 
\section{CONCLUSIONS}
\label{sec:conclusion}
This paper proposes an HMM-based data augmentation approach along with a language model for synthesizing conversational text. It is observed that the utterances synthesized using the proposed approach make fewer errors without compromising the quality of naturalness compared to those generated with a baseline E2E model. The novelty of the work comes from seamlessly combining the robustness of HTS and the synthesis quality of neural-network-based E2E models. Most importantly, we highlight the importance of the proposed method in a low-resource scenario where the state-of-the-art TTS models may not perform well. Since the proposed approach is architecture-agnostic, it can be further explored in the context of non-autoregressive models such as FastSpeech and FastSpeech2.

\bibliographystyle{IEEEtran}
\bibliography{IEEEexample}

\end{document}